\documentclass[aps,prl,twocolumn,amsmath,amssymb,amsfonts,nofootinbib,long,floatfix,showpacs]{revtex4}
\usepackage{epsfig,latexsym,bm,graphicx,epstopdf}

\newcommand\GeV{\mbox{GeV}}

\newcommand\Mpc{\mbox{Mpc}}

\newcommand\kk{\mathbf{k}}

\begin{document}

\title{Smooth magnetogenesis}

\author{L. Campanelli$^{1}$}
\email{leonardo.campanelli@ba.infn.it}

\author{A. Marrone$^{1,2}$}
\email{antonio.marrone@ba.infn.it}

\affiliation{$^1$Dipartimento di Fisica, Universit\`{a} di Bari, I-70126 Bari, Italy}
\affiliation{$^2$INFN - Sezione di Bari, I-70126 Bari, Italy}


\begin{abstract}
In the Ratra scenario of inflationary magnetogenesis,
the kinematic coupling between the photon and the inflaton undergoes a
nonanalytical jump at the end of inflation.
Using smooth interpolating analytical forms of the coupling function,
we show that such unphysical jump does not invalidate the main prediction of the model,
which still represents a viable mechanism for explaining cosmic magnetization.
Nevertheless, there is a spurious result associated with the nonanaliticity of the coupling,
to wit, the prediction that
the spectrum of created photons has a power-law decay in the
ultraviolet regime. This issue is discussed using both semiclassical approximation
and smooth coupling functions.
\end{abstract}




\maketitle


\section{I. Introduction}

The ubiquitous presence of large-scale cosmological magnetic fields is still un unexplained
feature of the Universe (for reviews on cosmic magnetic fields, see~\cite{Review1,Review2,Review3,Review4,Review5,Review6}).

One possibility is that such fields are relics from inflation, as first suggested by
Turner and Widrow long time ago~\cite{Turner-Widrow}. The idea that a spontaneous
inflationary magnetogenesis could be at the origin of cosmic magnetic fields
has been since then deeply investigated. After the seminal works by Ratra~\cite{Ratra}
and Dolgov~\cite{Dolgov}, numerous suggestions have been put forward
in an attempt to explain cosmic magnetization~\cite{G1,G2,G3,G4,G5,G6,G7,G8,G9,G10,G11,G12,G13,
G14,G15,G16,G17,G18,G19,G20,G21,G22,G23,G24,G25,G26,G27,G28,G29},
and to overcome a problem associated with inflationary magnetogenesis,
known as the ``strong coupling'' problem~\cite{Demozzi}.

Recently enough~\cite{Campanelli,Campanelli2}, we have reconsidered the Ratra model
and showed that it is a viable magnetogenesis mechanism able to explain
the presence of galactic and extragalactic magnetic fields
without suffering from the aforementioned problem.

A necessary condition to creating large-scale magnetic fields at inflation is to
break the conformal invariance of standard electromagnetism. 
In the Ratra model, this is attained by kinematically coupling the
photon field $A_\mu$ to the inflaton field $\phi$.
The modified electromagnetic Lagrangian turns to be~\cite{Ratra}
\begin{equation}
\label{L}
{\mathcal L}_{\rm Ratra} = -\frac14 f(\phi) F_{\mu\nu}F^{\mu\nu},
\end{equation}
where $F_{\mu\nu}$ is the electromagnetic field strength tensor and,
assuming homogeneity and isotropy, the coupling function $f(\phi)$
depends only on the (conformal) time. In real models of magnetogenesis,
such a coupling must ($i$) be a smooth function of time,
\footnote{Here, we are not considering any possible discontinuous
photon-inflaton coupling which could arise
if the inflaton undergoes a first-order phase transition during inflation and/or
reheating, namely we stay close to the original Ratra model (and its extensions) where
inflation and reheating are assumed to proceed smoothly without discontinuous jumps.}
($ii$)
approach unity after reheating, so to recover standard electromagnetism
in radiation and matter eras, and ($iii$) be greater than unity otherwise,
so to avoid the strong coupling problem.

In the original work by Ratra, and in its following extensions, however,
a simplified discontinuous form of $f(\phi)$ was assumed. This is because of a complete lack
of any theoretical and/or phenomenological particle physics model supporting the Ratra scenario,
which could have eventually justified and/or suggested a continuous form of the coupling.

In Sec.~IV, we argue, by using particular smooth forms of $f(\phi)$,
that the assumption of a discontinuous coupling
does not invalidate the main predictions of the model.
Needless to say, such a conclusion does not represent a general proof of this point,
proof that could eventually be given only if the exact form of the Ratra coupling were known.

Finally, we show that there are spurious results associated with a discontinuous coupling,
such as the fact that the spectrum of produced photons decays as a power law in the
ultraviolet regime, instead of falling off exponentially as
generally predicted by the standard theory of quantum theory in curved spacetime.
This latter feature of the Ratra model will be discussed in
Sec.~III by using the Wentzel-Kramers-Brillouin (WKB) approximation.
In the next section, we set our notations and we
briefly review the Ratra scenario in the light of the results obtained in~\cite{Campanelli}.

\section{II. Creation of magnetic fields in the Ratra model}

The creation of inflationary magnetic fields in the Ratra model is a pure
quantum effect~\cite{Campanelli}. It is understood as the creation of photons out of the vacuum by the
changing gravitational field between two temporal regions, the ``in-region'' and the ``out-region,''
possessing two different, inequivalent vacua.
For the sake of convenience, we may define these regions by
$\eta \rightarrow \mp \infty$, respectively, where $\eta$ is the conformal time.
Physically, the former represent the temporal region where inflation starts and the latter
any temporal region after reheating.

Working in Fourier space, let us rescale
the photon wave functions $A_{k}^{({\rm in,out})}$ in the in- and out-regions as
\begin{equation}
\label{psiA}
\psi_{k}^{({\rm in,out})} = \sqrt{f/2k} \, A_{k}^{({\rm in,out})}.
\end{equation}
The in- and out-$\psi$ modes reduce to plane waves in the corresponding in- and out regions, $\psi_{k}^{({\rm in,out})} = e^{-ik\eta}/\sqrt{2k}$ for $\eta \rightarrow \mp \infty$,
and satisfy the equation of motion~\cite{Campanelli}
\begin{equation}
\label{psi}
\ddot{\psi}_{k} = U_k {\psi}_{k}.
\end{equation}
Here,
\begin{equation}
\label{Uk}
U_k(\eta) = -k^2 + U_0(\eta),
\end{equation}
$k$ is the comoving wave number, and
\begin{equation}
\label{U0}
U_0(\eta) = \frac{\ddot{\!\!\!\!\sqrt{f}}}{\sqrt{f}}
\end{equation}
will be referred to as the ``Ratra potential.''
The in- and out-modes are not independent but related trough the Bogoliubov  transformation~\cite{Campanelli}
\begin{equation}
\label{in-out}
\psi_{k}^{({\rm in})} =
\alpha_{k} \psi_{k}^{({\rm out})} + \beta_{k} \psi_{k}^{({\rm out}) *},
\end{equation}
where the Bogoliubov coefficients $\alpha_{k}$ and $\beta_{k}$
satisfy the Bogoliubov condition
\begin{equation}
\label{condition}
|\alpha_{k}|^2 - |\beta_{k}|^2  = 1.
\end{equation}
The in-vacuum state will contain out-particles so long as $\beta_{k} \neq 0$.
It can be easily shown that the number of particles, in this case, is~\cite{Campanelli,Birrell-Davies}
\begin{equation}
\label{n}
n_{\kk} = |\beta_{k}|^2.
\end{equation}
For a conformal theory, as the standard Maxwell theory in a
Friedmann-Robertson-Walker spacetime, the in- and out-vacua are equivalent and one has $\beta_{k} = 0$,
so that no particle creation occurs (this is the content of the well-known ``Parker theorem''~\cite{Birrell-Davies}).

Neglecting any possible effect of magnetohydrodynamic
turbulence~\cite{MHD1,MHD2,MHD3,MHD4,MHD5,MHD6,MHD7,MHD8} after reheating,
the actual magnetic field on a scale $\lambda = 1/k$
is~\cite{Campanelli} (see also~\cite{Campanelli2,Tsagas})
\begin{equation}
\label{B0}
B_k \sim k^2 \sqrt{n_\kk} \, ,
\end{equation}
in the (quasiclassical) limit $n_\kk \gg 1$.
The actual magnetic field spectrum on large (cosmological) scales is then completely determined by
the density number of photons with wave number $k$ created at inflation.

In the (original) Ratra model, the coupling function is
\begin{equation}
\label{Ra1}
f(\eta) =
\left\{ \begin{array}{lll}
  f_*(\eta_e/\eta)^{2p} \, , & \eta \leq \eta_e, \\
  1,   & \eta > \eta_e,
  \end{array}
  \right.
\end{equation}
where $f_*>1$, $\eta_e$ is the conformal time at the end of inflation,
\footnote{Without loss of generality, we can assume
that $\eta_e$ is negative-defined as in the case of pure de Sitter inflation.}
and $p$ is a real parameter which we assume, for the sake of simplicity, to be less than $-1$. This, in turns, gives for the Ratra potential the simple expression
\begin{equation}
\label{Ra2}
U_0(\eta) =
\left\{ \begin{array}{lll}
  \frac{p(p+1)}{\eta^2}\, , & \eta \leq \eta_e, \\
  0,  & \eta > \eta_e.
  \end{array}
  \right.
\end{equation}
Modes well outside the horizon at the end of inflation, $-k\eta_e \ll 1$, are efficiently produced~\cite{Campanelli,Campanelli2},
\begin{equation}
\label{Ra12}
n_\kk \simeq \frac{[2^{-p}(p+1)\Gamma(-p-1/2)]^2}{16\pi} \:(-k\eta_e)^{2p},
\end{equation}
where $\Gamma(x)$ is the gamma function.
Conversely, the production of photons with wavelength much smaller than the horizon length at the end of inflation, $-k\eta_e \gg 1$, is suppressed as~\cite{Campanelli}
\begin{equation}
\label{Ra13}
n_\kk \simeq \left[\frac{p(p+1)}{4} \right]^{\!2} (-k\eta_e)^{-4}.
\end{equation}
Such an unphysical power-law decay of the photon spectrum in the ultraviolet regime
can be attributed to the nonanalyticity of the Ratra potential at the
point $\eta_e$. Indeed, for smooth potentials,
the general expectation is that of an exponential decay
(see Sec.~IIIb and the discussion at the beginning of Sec.~IV).

\section{III. Particle creation analogy with quantum potential scattering}

The equation of motion~(\ref{psi}) of the $\psi$ modes is formally equal to the
one-dimensional Schrodinger equation with zero energy and potential energy $U_{k}$, with
$\eta$ taking the place of the spatial coordinate and $k$ being an external parameter.
Continuing the analogy, Eq.~(\ref{in-out}) connecting the
$\psi_{k}^{(\rm in)}$ and $\psi_{k}^{(\rm out)}$ modes describes the
scattering of $\psi$-waves off the potential $U_{k}$, the incident, reflected, and transmitted
waves being $\alpha_{k} \psi_{k}^{(\rm out)}$, $\beta_{k} \psi_{k}^{(\rm out) *}$,
and $\alpha_{k} \psi_{k}^{(\rm in)}$,
respectively. This is schematically illustrated in Fig.~1.


\begin{figure*}[t!]
\begin{center}
\includegraphics[clip,width=0.7\textwidth]{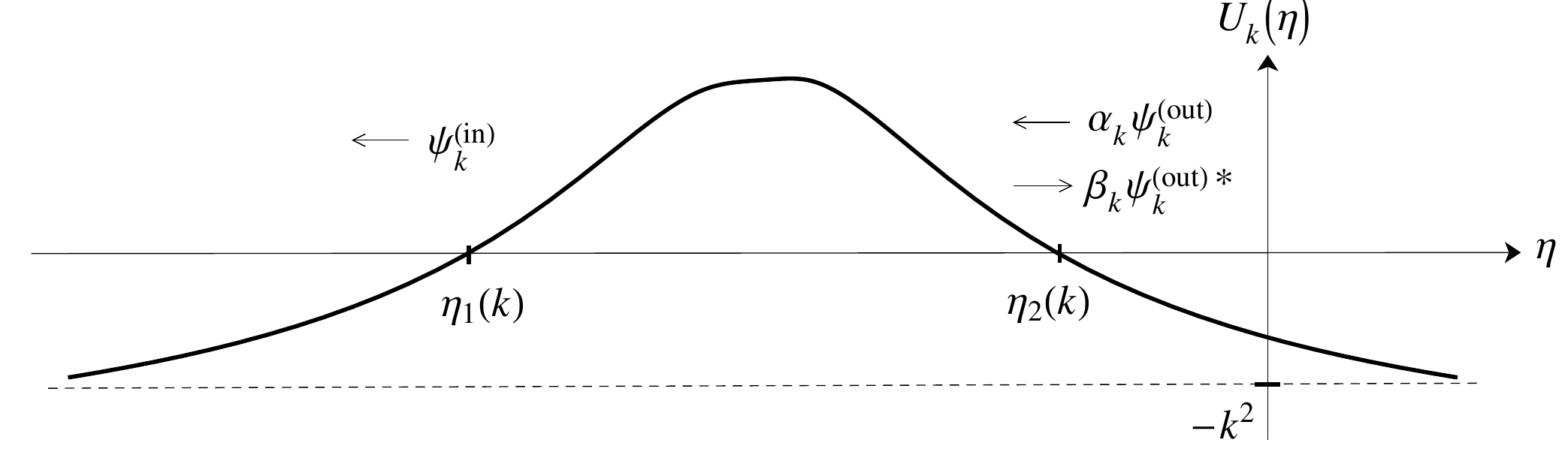}
\caption{Schematic illustration of the equivalence of the problem of potential scattering in
quantum mechanics and the creation of particles in a curved background.
The formal relations between the spectrum of particles created out from the vacuum $n_\kk$
and the reflection and transmission coefficients through the barrier $U_k(\eta)$ is
given in Eqs.~(\ref{Ref})-(\ref{Tr}). The points $\eta_{1,2}(k)$ are the classical
turning points for the quantum-mechanical problem of penetration through the potential barrier.}
\end{center}
\end{figure*}


Moreover, one can define a density current associated to any $\psi$-mode as
\begin{equation}
\label{current}
j_{k} = \langle \psi_{k} | \psi_{k}^* \rangle,
\end{equation}
where the inner product of any two solutions $\psi_{k}^{(1)}$ and $\psi_{k}^{(2)}$ of Eq.~(\ref{in-out}) is defined by~\cite{Campanelli}
\begin{equation}
\label{inner}
\langle \psi_{k}^{(1)} | \psi_{k}^{(2)} \rangle =
-i (\psi_{k}^{(1)} \dot{\psi}_{k}^{(2)} - \dot{\psi}_{k}^{(1)} \psi_{k}^{(2)}).
\end{equation}
The conservation of the current~(\ref{current}), $\dot{j}_{k} = 0$, follows directly from the conservation of the inner product.
The incident, reflected, and transmitted currents are then
\begin{eqnarray}
\label{currents}
&& j_{k}^{({\rm inc})} =
\langle \alpha_{k} \psi_{k}^{({\rm out})} | \alpha_{k}^* \psi_{k}^{({\rm out}) *} \rangle = |\alpha_{k}|^2, \\
&& j_{k}^{({\rm ref})} =
\langle \beta_{k} \psi_{k}^{({\rm out}) *} | \beta_{k}^* \psi_{k}^{({\rm out})} \rangle = -|\beta_{k}|^2, \\
&& j_{k}^{({\rm tr})} =
\langle \psi_{k}^{({\rm in})} | \psi_{k}^{({\rm in}) *} \rangle = 1,
\end{eqnarray}
where we used the fact that~\cite{Campanelli}
\begin{eqnarray}
\label{current1}
&& \alpha_{k} = \langle \psi_{k}^{({\rm in})} | \psi_{k}^{({\rm out}) *} \rangle, \\
\label{current2}
&& \beta_{k} = -\langle \psi_{k}^{({\rm in})} | \psi_{k}^{({\rm out})} \rangle,
\end{eqnarray}
and
$\langle \psi_{k}^{({\rm in})} | \psi_{k}^{({\rm in}) *} \rangle = \langle \psi_{k}^{({\rm out})} | \psi_{k}^{({\rm out}) *} \rangle = 1$.

Taking into account Eq.~(\ref{n}) and the Bogoliubov condition~(\ref{condition}), we find
the reflection and transmission coefficients
\begin{eqnarray}
\label{Ref}
&& R_{k} = -\frac{j_{k}^{({\rm ref})}}{j_{k}^{({\rm inc})}} = \frac{n_{\kk}}{1+n_{\kk}} \, , \\
\label{Tr}
&& T_{k} = \frac{j_{k}^{({\rm tr})}}{j_{k}^{({\rm inc})}} = \frac{1}{1+n_{\kk}} \, ,
\end{eqnarray}
from which the unitarity condition $R_{k} + T_{k} = 1$ directly follows.

Let us assume, for the sake of simplicity, that the Ratra potential $U_0(\eta)$
is positive and vanishes in the in- and out regions $\eta \rightarrow \mp \infty$.
It is clear that if $k^2 < \max U_0(\eta)$, then the ``particle''
described by the wave function $\alpha_{k} \psi_{k}^{(\rm out)}$ will
penetrate through the potential barrier $U_k(\eta)$.
To ``particles'' which deeply penetrate into the barrier, $k^2 \ll \max U_0(\eta)$,
there will correspond a large reflection coefficient and, in turn,
by Eq.~(\ref{Ref}), a large particle number $n_\kk$.
On the other hand, if $k^2 > \max U_0(\eta)$, the
``particle'' is reflected above the barrier. For $k^2 \gg \max U_0(\eta)$,
the reflection coefficient for scattering above the barrier will be small.
To this case, there will correspond a small production of particles, $n_\kk \ll 1$.

The usefulness of Eqs.~(\ref{Ref})-(\ref{Tr}) resides in the fact that if the
potential $U_k$ is slowly varying, in the sense specified below,
one can apply the standard semiclassical (WKB) results for the
reflection and transmission coefficients. Using the formal equivalence
of the two problems of potential scattering in
quantum mechanics and the creation of
particle out from the vacuum in a curved spacetime, one can then
find the expression for the particle number $n_\kk$.

\subsection{IIIa. Particle number in WKB approximation}

The WKB approximation is valid whenever
the potential $U_k$ satisfies the semiclassical condition~\cite{Landau}
\begin{equation}
\label{sm1}
\left|\frac{\dot{U}_k}{2U_k^{3/2}} \right| \ll 1.
\end{equation}
If this is the case, the reflection coefficient is either small or large,
and the reflection takes place either deeply through the barrier or
above it. Therefore, according to the above discussion, the WKB approximation will
provide us with the interesting cases of large and small $n_\kk$.

{\it Large particle number.} -- Let us assume, for the sake of simplicity, that
the barrier has a bell shape form, as in Fig.~1. Accordingly,
there will be two classical turning points, $\eta_{1}(k) < \eta_{2}(k)$,
for a deep penetration through the potential barrier.
Since in this case $R_k \simeq 1$, and then $T_k \ll 1$, we have from Eq.(\ref{Tr}),
$n_\kk \simeq T_k^{-1} \gg 1$. Using the standard result for the
expression of the transmission coefficient in WKB approximation~\cite{Landau},
we find
\begin{equation}
\label{sm2}
n_\kk \simeq e^{2S_k},
\end{equation}
where
\footnote{As it straightforward to check, the quantity $S_k$ gives, in the out-region, the so-called
``squeezing'' parameter $r_k(\eta)$ defined in~\cite{Campanelli},
$\lim_{\eta \rightarrow +\infty} r_k(\eta) = S_k + \ln 2$. Large values of the squeezing parameter
indicate that the system (the magnetic field) is in a (quasi-)classical state.}
\begin{equation}
\label{sm3}
S_k = \int_{\eta_1(k)}^{\eta_2(k)} \! d\eta \sqrt{U_k(\eta)} \, .
\end{equation}
The magnetic field spectrum is proportional to the
square root of the particle number [see Eq.~(\ref{B0})], the latter
representing an amplification factor of the in-vacuum magnetic fluctuations. As first noticed by
Giovannini~\cite{Review2}, the amount of such an amplification depends,
for a given mode, on the ``time spent'' under the potential barrier.
From Eqs.~(\ref{sm2})-(\ref{sm3}), we see that this ``time'' is indeed the area enclosed by the
square root of the potential between the two classical turning points.

{\it Small particle number.} -- The probability that a ``particle'' is scattered above the
potential barrier is small for large values of $k^2$ compared
to the height of the Ratra barrier. Using Eq.~(\ref{Ref}), we
then have $n_\kk \simeq R_k \ll 1$. Using the standard result for the
expression of the reflection coefficient in WKB approximation~\cite{Landau},
we find
\begin{equation}
\label{sm4}
n_\kk \simeq e^{-4 \sigma_k},
\end{equation}
where
\begin{equation}
\label{sm5}
\sigma_k = \mbox{Im} \int_{\eta_R}^{\eta_I(k)} \! d\eta \sqrt{-U_k(\eta)} \, .
\end{equation}
Here, $\eta_I(k)$ is the so-called imaginary turning point,
the complex solution of the equation $U_k(\eta) = 0$ for $k^2 > \max U_0(\eta)$,
and $\eta_R$ is an arbitrary and inessential real parameter.
The integration in Eq.~(\ref{sm5}) has to be performed in the complex upper half-plane
($\mbox{Im}[\eta_I(k)] > 0$).
If the equation for the imaginary turning point admits more than one solution,
one must select the one for which $\sigma_k$ is smallest~\cite{Landau}.

Let us finally observe that, for $k \rightarrow \infty$
[namely for sufficiently large wave numbers, $k^2 \gg \max(U_0,|\dot{U}_0|^{2/3})$],
condition~(\ref{sm1}) is satisfied
whatever is the form of the Ratra potential and the WKB approximation
is then applicable. Moreover, if there exists the limit
$\lim_{k \rightarrow \infty} \eta_I(k) = \eta_I(\infty)$, Eq.~(\ref{sm4}) takes the simple form
\begin{equation}
\label{sm6}
n_\kk \simeq  e^{-4k \, \mathrm{Im} [\eta_I(\infty)]},
\end{equation}
at the leading order in $k \rightarrow \infty$. In the case of multiple
imaginary turning points, one has to take the one which has the smallest (positive)
imaginary part.

\subsection{IIIb. Ratra model in WKB approximation}

As an application of the above results, let us consider in
semiclassical approximation the exactly solvable Ratra model defined
by the potential~(\ref{Ra2}). The WKB applicability condition~(\ref{sm1}) in this case reads
\begin{equation}
\label{exampleU}
q \left| 1 - (k\eta/q)^2 \right |^{3/2} \gg 1,
\end{equation}
where we have defined $q = \sqrt{p(p+1)}$.
The classical turning points are easily found,
\begin{equation}
\label{eta12}
\eta_1(k) = -\frac{q}{k} \, , \;\;\; \eta_2(k) = \eta_e,
\end{equation}
for $-k\eta_e < q$.
At large scales, $-k\eta \ll q$, far from
from the turning points (where the WKB approximation cannot be trustful),
condition~(\ref{exampleU}) is satisfied for $q \gg 1$, namely for $p \ll -1$.
Using Eqs.~(\ref{sm2})-(\ref{sm3}), we find
\begin{equation}
\label{example1}
n_\kk^{\rm (WKB)} = (2q/e)^{2q} (-k\eta_e)^{-2q}.
\end{equation}
In the limit $p \rightarrow -\infty$, this approximate result
and the exact result~(\ref{Ra12}) coincide.

However, the WKB expression~(\ref{example1})
is not a good approximation for values of $|p|$ of order unity.
\footnote{In order to avoid the so-called backreaction problem
of inflationary magnetogenesis~\cite{Demozzi}, one
must take $|p| \leq 2$~\cite{Campanelli}.}
This is due to the fact that $-k\eta_e$ is a very small quantity, so that even a
small change in the value of its exponent produces a large deviation of the particle
number and, in turn, of the actual magnetic field.
To see this, let us use Eq.~(\ref{B0}) and
take into account that~\cite{Campanelli2}
\begin{equation}
\label{keta}
-k\eta_e \simeq 10^{-21} \! \left(\frac{M}{10^{16}\GeV}\right)^{\!\!-2/3} \!\!
\left(\frac{T_{\rm RH}}{10^{10}\GeV}\right)^{\!\!-1/3} \! \frac{k}{\Mpc^{-1}} \, ,
\end{equation}
where
$M$ is the scale of inflation and $T_{\rm RH}$ the reheat temperature.
We have $B_k^{\rm (WKB)} / B_k \sim (-k\eta_e)^{q-|p|}$ which, for example, results in
$B_k^{\rm (WKB)} / B_k \sim 10^{12}$ for $p=-2$,
$M = 10^{16}\GeV$, $T_{\rm RH} = 10^{10}\GeV$, and $\lambda = 1/k = \Mpc$.

At small scales, $-k\eta \gg q$, the WKB approximation is valid.
Nevertheless, we cannot apply Eqs.~(\ref{sm4})-(\ref{sm5}) in the model at hand, since
they are valid only for smooth potentials $U_k(\eta)$.
If there is a discontinuity in the Ratra potential $U_0(\eta)$, as in Eq.~(\ref{U0}),
the reflection coefficient is determined mainly by the wave function at that point~\cite{Landau}.
In this case, the perturbation theory can be used to calculate $R_k$ in the case of a
quasiclassical barrier. Using the standard results in~\cite{Landau}, we find that
\begin{equation}
\label{Landau}
n_\kk \simeq R_k \simeq \frac{|U_0(\eta_\star)|^2}{16k^4} \, ,
\end{equation}
where $\eta_\star$ is the discontinuity point. Applying this result to the Ratra model,
the discontinuity point being $\eta_e$, we find that the perturbation
theory gives the exact result for $n_\kk$ in the limit $-k\eta_e \gg q$, namely Eq.~(\ref{Ra13}).

We stress the fact that the perturbation theory is always valid for sufficiently large wave numbers (as already observed above). Hence, any smooth form of Ratra coupling $f(\eta)$ would lead to an exponential decay of the particle number in the ultraviolet regime [see Eq.~(\ref{sm6}], in contrast to the $k^{-4}$ law predicted by the discontinuous coupling.

\section{IV. Smooth coupling in the Ratra model}

As already noticed, the coupling $f(\eta)$
must be a smooth function
of time in any real model of inflationary magnetogenesis. If, moreover, such a coupling function
is slowly varying in the in- and out-regions, in the sense that
\begin{equation}
\label{exp1}
\lim_{\eta \rightarrow \pm \infty} \frac{d^n}{d\eta^n} \, \frac{\dot{f}}{f} = 0,
\end{equation}
for all $n \in \mathbb{N}$, then the in- and out- electromagnetic vacuum states are vacua of infinite
adiabatic order~\cite{Campanelli}.
\footnote{It is also assumed that the (spatially flat, Friedmann-Robertson-Walker) spacetime
is slowly varying in the in- and out-regions,
$\lim_{\eta \rightarrow \pm \infty} \frac{d^n}{d\eta^n} \, \frac{\dot{a}}{a} = 0$,
where $a(\eta)$ is the expansion parameter.
This condition is automatically satisfied in (de Sitter or power-law) inflation
as well as in radiation and matter dominated eras.}
This, in turns, implies that the number of
produced particle must go to zero exponentially or, more precisely, faster than any power of
$k$ in the limit $k \rightarrow +\infty$~\cite{Birrell-Davies},
\begin{equation}
\label{exp2}
\lim_{k \rightarrow \infty} \frac{n_\kk}{k^{n+1}} = 0,
\end{equation}
for all $n \in \mathbb{N}$. In this section, we consider examples of
coupling functions which smoothly interpolate the nonanalytical profile $f(\eta)$
of the original Ratra model.
We will see that an exponential decay of the particle number at
large wave numbers will replace
the (nonphysical) power-law decay discussed in Secs.~II and IIIb.
At small wave numbers, instead, all the results obtained in the
discontinuous case will be confirmed. This is to be expected,
since long wavelength modes should not be strongly
affected by variations of the coupling function on small (time)scales.

\subsection{IVa. Analytical example}

As a first example, let us consider the following coupling,
\begin{equation}
\label{EE2}
f(\eta) = (1 + e^{\eta/\eta_*})^{2b} \, [_2F_1(b,b,1,-e^{\eta/\eta_*})]^2,
\end{equation}
where $\eta_* < 0$ is a free parameter (that could represent, for example,
the time at the end of inflation $\eta_e$), $_2F_1(a,b,c,z)$ is the Gauss hypergeometric function~\cite{Abramowitz},
and we assume that the positive parameter $b$ is in the range $b \in [1/2,1]$.
A plot of the coupling function, for two different values of $b$, is given in Fig.~2.
The asymptotic expansions of $f(\eta)$ are
\begin{equation}
\label{EE2bis}
f(\eta) =
\left\{ \begin{array}{lll}
   \left( \frac{\sin(\pi b) \frac{\eta}{\eta_*} + \gamma_b}{\pi} \right)^2 + \mathcal{O}(e^{-\eta/\eta_*}), & \eta \rightarrow -\infty , \\ \\
   1 + 2b (1-b) \, e^{\eta/\eta_*} + \mathcal{O}(e^{2\eta/\eta_*}), & \eta \rightarrow +\infty,
  \end{array}
  \right.
\end{equation}
where $\gamma_b = - \sin(\pi b) [\gamma + H_{-b} + \psi(b)]$ is a real increasing function of $b$
such that $\gamma_{1/2} = 4\ln2$ and $\gamma_1 = \pi$. Here, $\gamma$ is the Euler's constant, $H_n$ is the harmonic number of order $n$, and $\psi(x)$ is the digamma function~\cite{Abramowitz}.
Therefore, $f(\eta)$ smoothly interpolates between the behaviors $f(\eta) \propto \eta^2$
for $\eta \ll \eta_*$ and $f(\eta) = 1$ for $\eta \gg \eta_*$ (see Fig.~2).
This analytical case, then, represents a smooth interpolation of the Ratra discontinuous power-law case
for $p \simeq -1$. Accordingly, we expect for the particle number the law $n_\kk \sim  k^{-2}$ for small values of $k$ and an ``exponential'' decay for large wave numbers.


\begin{figure}[t!]
\begin{center}
\includegraphics[clip,width=0.45\textwidth]{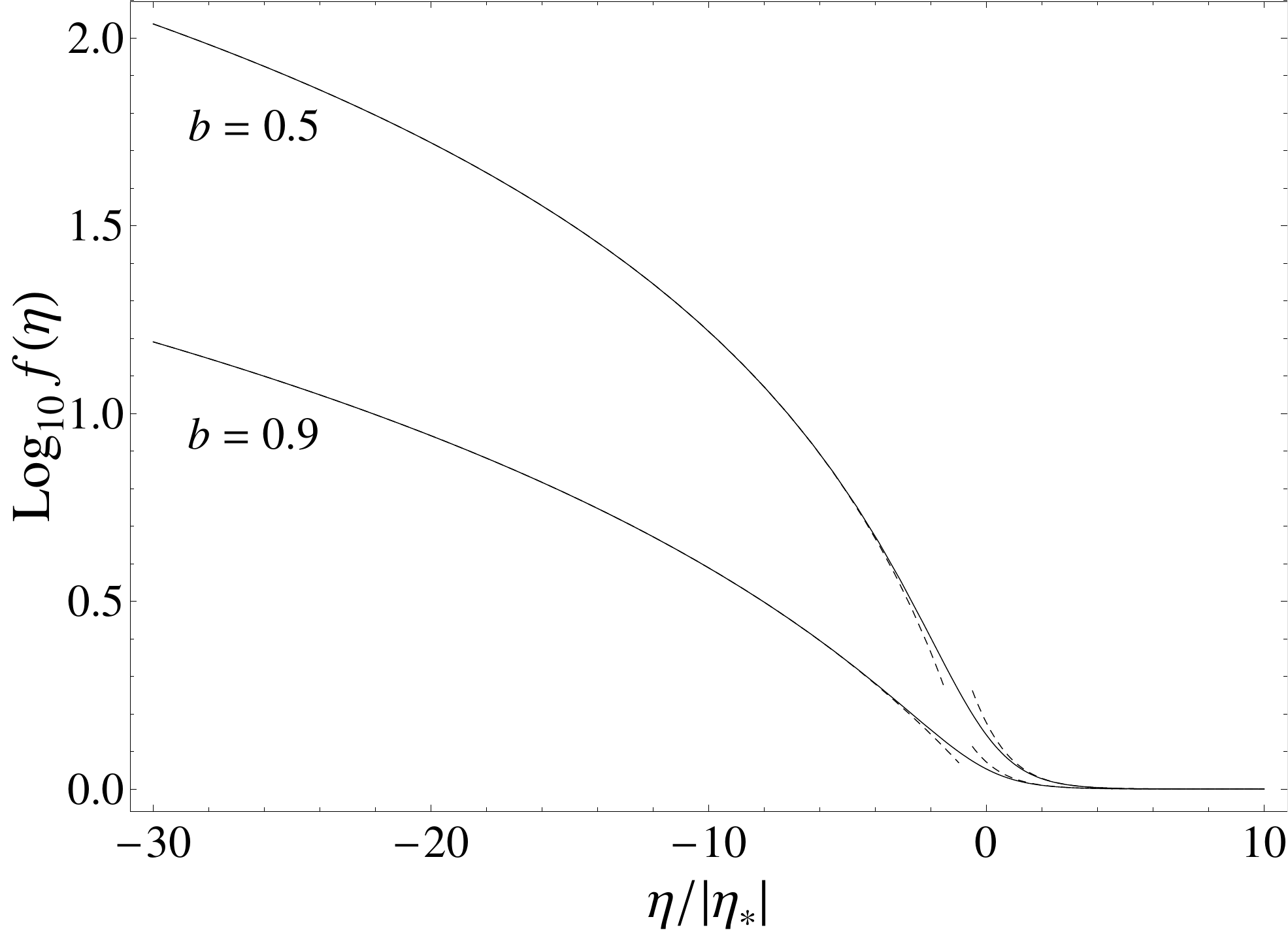}
\caption{The smooth coupling $f(\eta)$ in Eq.~(\ref{EE2}) as a function
of the conformal time for two different values of $b$. Dashed lines are the corresponding asymptotic expansions in Eq.~(\ref{EE2bis}).}
\end{center}
\end{figure}


The shape of the Ratra potential follows from Eq.~(\ref{EE2}),
\begin{equation}
\label{EE1}
U_0(\eta) = \frac{k_*^2}{1+\cosh (\eta/\eta_*)} \, ,
\end{equation}
where we have defined $k_* = \sqrt{\frac12 b(1-b)} / |\eta_*|$.

A standard calculation gives, for the Bunch-Davies-normalized out-$\psi$ modes,
\begin{eqnarray}
\label{EE5}
\!\!\!\!\!\!\!\!\!\!\!\!
\psi_k^{(\rm out)}(\eta) \!\!& = &\!\! \frac{e^{-ik\eta}}{\sqrt{2k}} \, (1 + e^{\eta/\eta_*})^{b} \nonumber \\
\!\!\!\!\!\!\!\!\!\!\!\!
\!\!& \times &\!\! _2F_1(b,b-2ik\eta_*,1-2ik\eta_*,-e^{\eta/\eta_*})
\end{eqnarray}
and, for the Bunch-Davies-normalized in-$\psi$ modes, the expression in Eq.~(\ref{in-out}) with
\begin{eqnarray}
\label{EE4a}
\alpha_k \!\!& = &\!\! \frac{\cosh(4\pi k\eta_*) - \cos(2\pi b)}{2\pi^2} \, \Gamma(2ik\eta_*) \Gamma(1 + 2ik\eta_*) \nonumber \\
          \!\!& \times &\!\! \Gamma(b - 2ik\eta_*) \Gamma(1 - b - 2ik\eta_*), \\
\label{EE4b}
\beta_k \!\!& = &\!\! i \, \frac{\sin(\pi b)}{\sinh(2\pi k \eta_*)} \, .
\end{eqnarray}
Equations~(\ref{EE4a}) and (\ref{EE4b}) are precisely the Bogoliubov coefficients defining the
in-$\psi$ modes in terms of the out-$\psi$ modes [a straightforward calculation
shows that the Bogoliubov condition~(\ref{condition}) is satisfied].
Taking the square modulus of $\beta_k$ we obtain the spectrum of the photons created out from the vacuum,
\begin{equation}
\label{EE8}
n_\kk = \frac{\sin^2(\pi b)}{\sinh^2(2\pi k \eta_*)} \, .
\end{equation}
For $b=1$ we have $k_* = 0$ so that
$U_0(\eta) = 0$, $f(\eta) = 1$, $\psi_k^{(\rm in,out)}(\eta) = e^{-ik\eta}/\sqrt{2k}$,
and in turn $n_\kk = 0$.

For small and large wave numbers, $-k\eta_* \ll 1$ and $-k\eta_* \gg 1$, respectively, we find
$n_\kk \sim  (-k\eta_*)^{-2}$ and $n_\kk \sim  e^{-4\pi k |\eta_*|}$, as we expected.

{\it WKB approximation.} --  Let us now apply the WKB approximation, which is always valid for
sufficiently large wave numbers, to the case under consideration.
The imaginary turning points in the complex upper half-plane for the potential~(\ref{EE1}) are
easily found,
\begin{equation}
\label{turning}
\eta_I(k) = |\eta_*|
\! \left[ \mbox{arccosh} \! \left( \frac{k_*^2}{k^2} - 1 \right) + 2 \pi i n \right] \!,
\end{equation}
for $k > k_*/\sqrt{2}$ and $n \in \mathbb{N}$. In the limit $k \rightarrow \infty$, we have
$\eta_I(\infty) = \pi i (1+2n) |\eta_*|$.
According to the general prescription discussed in Sec.~IIIa, we must take $n=0$, so that,
using Eq.~(\ref{sm6}), we obtain
\begin{equation}
\label{WKBN2}
n_\kk^{\rm (WKB)} \simeq  e^{-4\pi k |\eta_*|},
\end{equation}
at the leading order in $k/k_* \sim -k \eta_* \gg 1$. The above WKB result is in perfect agreement
with the exact result~(\ref{EE8}) in the large-wavenumber regime.

%

\subsection{IVb. Numerical example}

The analytical case discussed in the previous section
gives an example of smooth interpolation of the Ratra discontinuous power-law case
$p \simeq -1$.
The most important case, that of scaling-invariant actual magnetic fields,
corresponds, however, to case $p=-2$ (see~\cite{Campanelli}).
We then generalize the previous example by considering the coupling function
\begin{equation}
\label{NE1}
f(\eta) = \left[ \frac{2}{\pi} \, \sqrt{1 + e^{\eta/\eta_*}} \, K(-e^{\eta/\eta_*}) \right]^{-2p} \!,
\end{equation}
where, as before, $\eta_* < 0$ is a free parameter, and $K(m)$ is the complete elliptic integral of the first kind~\cite{Abramowitz}.
Since
\begin{equation}
\label{NE2}
f(\eta) =
\left\{ \begin{array}{lll}
   \left( \frac{\frac{\eta}{\eta_*} + \ln 16}{\pi} \right)^{\!-2p} + \mathcal{O}(e^{-\eta/\eta_*}), & \eta \rightarrow -\infty , \\ \\
   1 -\frac{p}{2} \, e^{\eta/\eta_*} + \mathcal{O}(e^{2\eta/\eta_*}), & \eta \rightarrow +\infty ,
  \end{array}
  \right.
\end{equation}
%
$f(\eta)$ smoothly interpolates between the behaviors $f(\eta) \propto \eta^{-2p}$
for $\eta \ll \eta_*$ and $f(\eta) = 1$ for $\eta \gg \eta_*$ (see Fig.~3).


\begin{figure}[t!]
\begin{center}
\includegraphics[clip,width=0.45\textwidth]{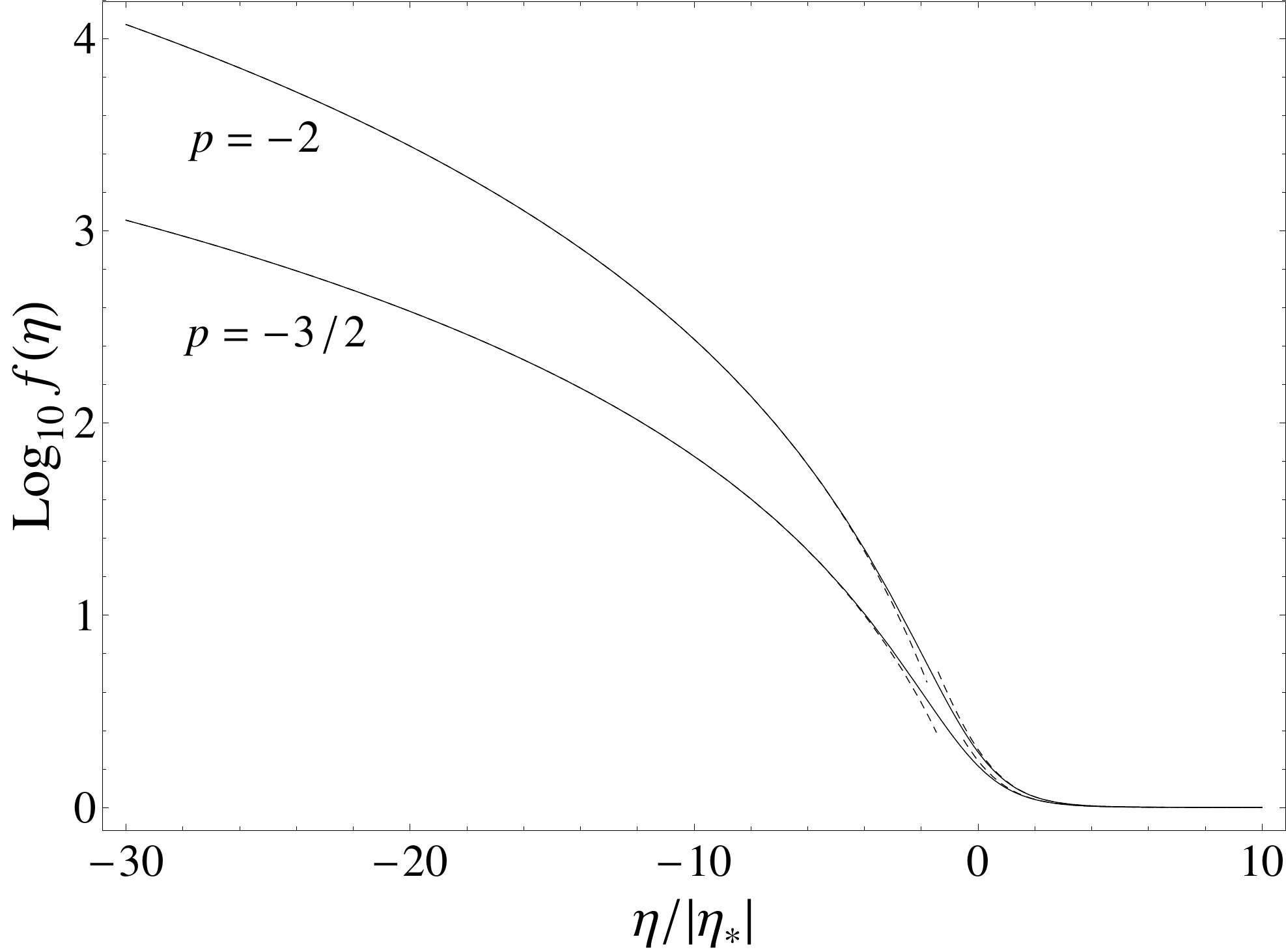}
\caption{The smooth coupling $f(\eta)$ in Eq.~(\ref{NE1}) as a function
of the conformal time for two different values of $p$. Dashed lines are the corresponding asymptotic expansions in Eq.~(\ref{NE2}).}
\end{center}
\end{figure}


The Ratra potential $U_0(\eta)$ reads
\begin{equation}
\label{NE3}
U_0(\eta) = \frac{k_*^2}{1+\cosh(\eta/\eta_*)} \, u_p(\eta),
\end{equation}
where
\begin{equation}
\label{NE3bis}
u_p(\eta)  = 1 - (p+1) \, e^{-\eta/\eta_*} \!
\left[1 - \frac{E(-e^{\eta/\eta_*})}{K(-e^{\eta/\eta_*})} \right]^2 \!,
\end{equation}
$k_* = \sqrt{\frac{|p|}{8}} / |\eta_*|$, and $E(m)$ is the complete elliptic integral of the second kind~\cite{Abramowitz}. The case $p=-1$ reduces to the (analytically solvable) case (with $b=1/2$) of Sec.~IVa.
The asymptotic expansions of the Ratra potential are
\begin{equation}
\label{NE4}
U_0(\eta) =
\left\{ \begin{array}{lll}
   \frac{p(p+1)}{\left(\eta + \eta_* \! \ln 16 \right)^2}  + \mathcal{O}(e^{-\eta/\eta_*}), & \eta \rightarrow -\infty , \\ \\
   -\frac{p}{4\eta_*^2} \, e^{\eta/\eta_*} + \mathcal{O}(e^{\hat{s}\eta/\eta_*}), & \eta \rightarrow +\infty .
  \end{array}
  \right.
\end{equation}
%
%
The equation of motion~(\ref{psi}) for the $\psi$ modes cannot be integrated
analytically for the general Ratra potential~(\ref{NE3}).
We then numerically integrate it by imposing the
Bunch-Davies boundary condition that
the in-$\psi$ mode reduces to the plane-wave solution $e^{-ik\eta}/\sqrt{2k}$ in the in-region $\eta \rightarrow - \infty$. In the same way, we may impose the Bunch-Davies boundary condition
in the out-region $\eta \rightarrow +\infty$ to find the out-$\psi$ mode
and then, using Eq.~(\ref{in-out}), we may obtain the Bogoliubov coefficients.
However, it is easier from a numerical point of view to calculate
the Bogoliubov coefficients starting from the expressions
\begin{eqnarray}
\label{alphaAlone1}
&& \alpha_{k} = \lim_{\eta \rightarrow + \infty} \frac{e^{ik\eta}}{\sqrt{2k}}
\left( k + i \frac{\partial }{\partial \eta}\right) \! \psi_{k}^{({\rm in})}, \\
\label{betaAlone1}
&& \beta_{k} = \lim_{\eta \rightarrow + \infty} \frac{e^{-ik\eta}}{\sqrt{2k}}
\left( k - i \frac{\partial }{\partial \eta}\right) \! \psi_{k}^{({\rm in})},
\end{eqnarray}
which make use only of the in-$\psi$ modes. Equations~(\ref{alphaAlone1}) and (\ref{betaAlone1})
come from Eqs.~(\ref{current1}) and (\ref{current2}), respectively, taking the limit
$\eta \rightarrow +\infty$ and observing that in the out-region
$\psi_{k}^{({\rm out})} = e^{-ik\eta}/\sqrt{2k}$.


\begin{figure}[t!]
\begin{center}
\includegraphics[clip,width=0.463\textwidth]{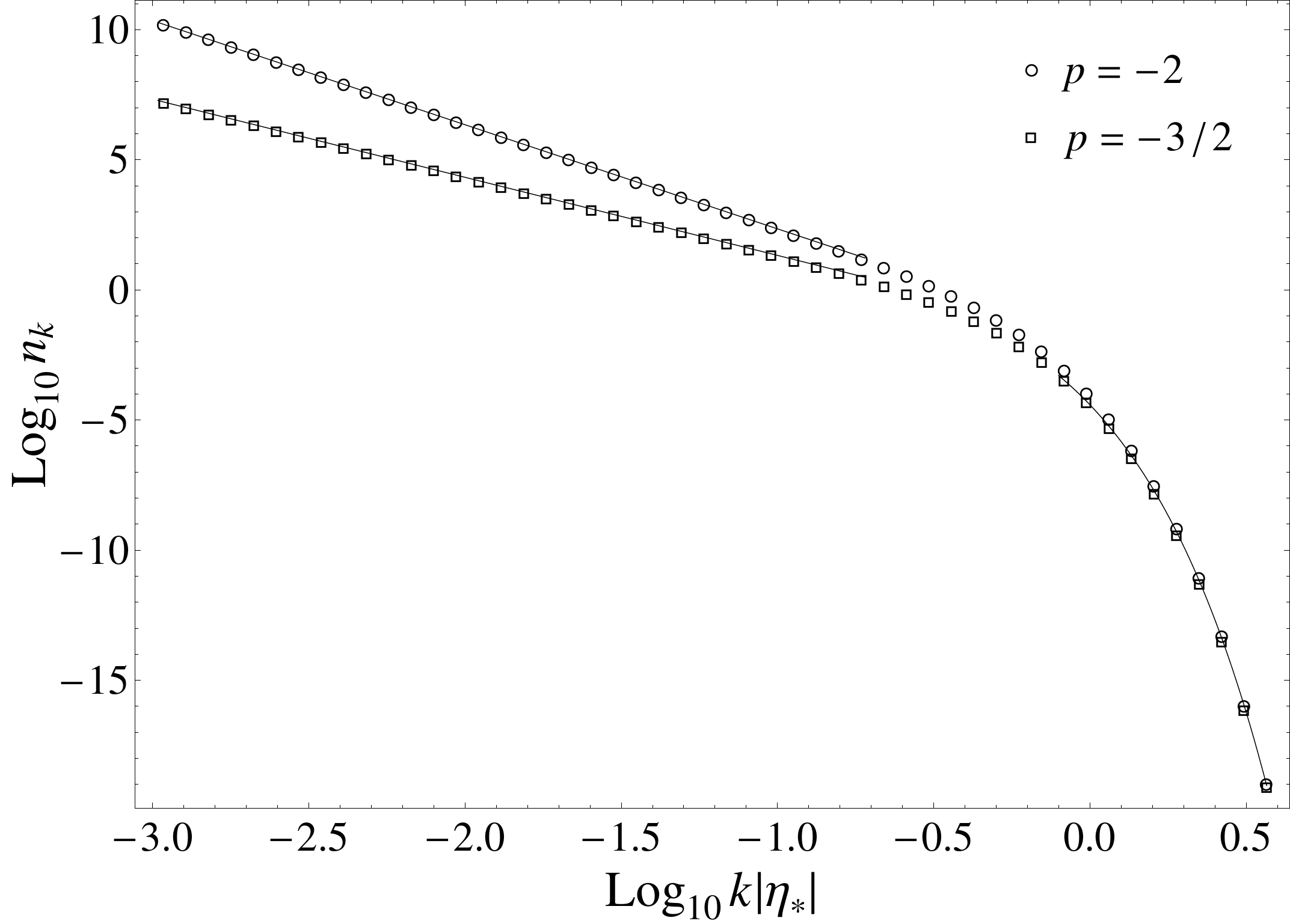}
\caption{The particle number $n_\kk$ as a function of $k|\eta_*|$ in the Ratra model described
by the coupling function~(\ref{NE1}). Continuous lines refer to the asymptotic numerical solutions described in the text [see Eqs.~(\ref{approx1}), (\ref{approx2}), and (\ref{approx3})].}
\end{center}
\end{figure}


In Fig.~4, we show the particle number $n_\kk = |\beta_k|^2$ for two different values of $p$ [we checked that the Bogoliubov condition~(\ref{condition}) holds]. Continuous lines in the small wave number regime ($-k\eta_* \ll 1$) correspond to
\begin{eqnarray}
\label{approx1}
&& n_\kk \simeq 0.021 \, (-k\eta_*)^{-3}, \\  
\label{approx2}
&& n_\kk \simeq 0.022 \, (-k\eta_*)^{-4},  
\end{eqnarray}
for $p=-3/2$ and $p=-2$, respectively. The exponents of these power laws agree with the results obtained in Ratra discontinuous model [see, in particular, Eq.~(\ref{Ra12})].
For large wave numbers ($-k\eta_* \gg 1$), instead, the continuous line is
\begin{equation}
\label{approx3}
n_\kk \simeq 11 \, e^{-4\pi k |\eta_*|}.
\end{equation}
Such an exponential decay of the particle number agrees well with the WKB solution discussed below.

{\it WKB approximation.} -- The imaginary turning points
$\eta_I(\infty) = \pi i (1+2n) |\eta_*|$ ($n \in \mathbb{N}$) of the Ratra
potential~(\ref{EE1}) are also imaginary turning points of the
potential~(\ref{NE3}) for $p\neq -2$.
This is because the function $u_p(\eta)$ tends to a nonvanishing constant when
$\eta$ approaches $\eta_I(\infty)$,
$\lim_{\eta \rightarrow \eta_I(\infty)} u_p(\eta) = 2 + p$.
In the case $p=-2$, instead, such a constant is zero and one needs
the next-to-leading order term in the expansion of the imaginary turning points
of~(\ref{EE1}),
$\eta_I(k) \simeq \eta_I(\infty) - i\sqrt{2}k_*/k$.
Inserting the above expression in Eq.~(\ref{NE3bis}), we find
that $u_{-2}$ vanishes logarithmically as $u_{-2}(\eta_I(k)) \simeq 8/\ln(k/k_*)$
in the limit $k \rightarrow \infty$.
The function which multiplies $u_p(\eta)$ in Eq.~(\ref{NE3}), instead,
diverges quadratically in $k$ in the same limit.
Accordingly, the
imaginary turning points of the Ratra potentials~(\ref{EE1})
and (\ref{NE3}) coincide in the limit $k \rightarrow \infty$
also in the case $p=-2$.

This result is confirmed by a direct numerical integration of the
equation defining the imaginary turning points of the
Ratra potential~(\ref{NE3}), $U_0(\eta_I(k)) = k^2$.

In Fig.~5, we show the imaginary part of the imaginary turning point
(with smallest positive imaginary part)
for two different values of $p$.
For large values of $k$, the imaginary part of $\eta_I(k)$ approaches the value
\begin{equation}
\label{Numeta}
\mbox{Im}[\eta_I(\infty)] \simeq \pi |\eta_*|.
\end{equation}
Using Eq.~(\ref{sm6}), then, we find that the particle number in WKB approximation is given by
Eq.~(\ref{WKBN2}) in the limit $k/k_* \sim -k\eta_* \gg 1$, as in the (analytical) case of Sec.~IVa.


\begin{figure}[t!]
\begin{center}
\includegraphics[clip,width=0.45\textwidth]{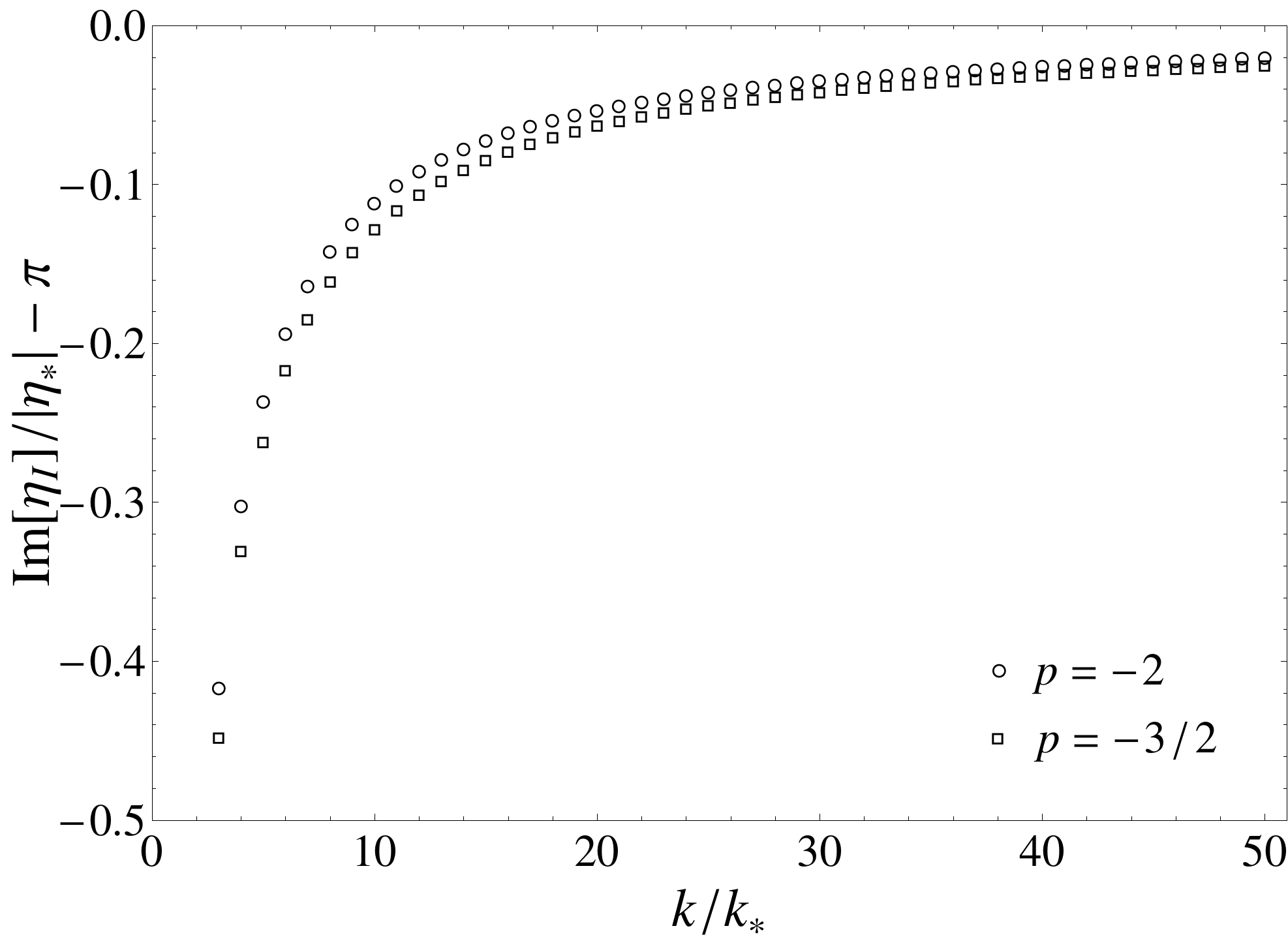}
\caption{Imaginary part of the imaginary turning point $\eta_I(k)$ as a function of the
wave number $k$ in the model described by the Ratra potential~(\ref{NE3}) for two different
values of $p$.}
\end{center}
\end{figure}


\section{V. Concluding remarks}

The Ratra model for the creation of magnetic fields during inflation has
received much interest in the last years. This is primarily due to its simplicity and to
its potential feature of explaining the origin of the observed galactic and extragalactic
large-scale magnetic fields.

In the original model, as well as in all subsequent variants of it,
the form of coupling function between the photon and the inflaton
plays the key role in the genesis of cosmological magnetic fields.
Such a nonstandard time-dependent coupling, which is responsible for the breaking
of the electromagnetic conformal invariance, has been always assumed to instantaneously
reduce to its standard constant value in radiation era. Such a nonanalytical
coupling could arise only in particular inflationary models,
where the inflaton, or other background fields coupled to it, evolve
discontinuously, as for example in a first-order phase transition.
However, this in not the case for the Ratra and Ratra-like models, since
they are constructed and framed in a smoothly evolving
inflationary and reheated universe.

In this paper, we have investigated the consequences of taking an analytical
coupling function that smoothly interpolates the (unphysical) discontinuous
coupling usually used in the Ratra and Ratra-like models.
Both using particular smooth forms of the coupling and working in
WKB approximation, we have studied the behavior of the
inflation-produced photon spectrum at small and large scales.
We have found an exponential decay in the ultraviolet part of such a spectrum,
in contrast with the (unphysical) power-law fall-off predicted
in the discontinuous case.
At large cosmological scales, however, the main results of the model remain unchanged
in the considered cases.
Since only these scales are important for cosmic magnetic fields, we conclude
that the Ratra model is still a viable mechanism of inflationary magnetogenesis.

Nevertheless, it should be said that the form of the coupling function used in this investigation
as well as those assumed in both the original work by Ratra and in its developments
do not have any solid theoretical and/or phenomenological justification.
Hence, as in the case of (scalar field) inflation, the Ratra mechanism is just a paradigm, although very elegant and attractive, in search of a background particle physics model.


\end{document}